\documentclass[12pt]{elsarticle}
\usepackage[dvipsnames]{xcolor}
\RequirePackage[T1]{fontenc}
\RequirePackage{mathptmx}
\usepackage{fullpage}
\usepackage{natbib}
\usepackage{slashed}
\usepackage{graphics,bm}
\usepackage{epsfig}
\usepackage{graphicx}
\usepackage{amsmath}
\usepackage{amssymb}
\usepackage{bbold}
\usepackage{wrapfig}
\usepackage{color}
\usepackage{hyperref}
\usepackage{mathtools}
\usepackage{cancel}
\usepackage{multirow}
\usepackage{upgreek}
\usepackage{subcaption}
    \usepackage[vertfit]{breakurl} 

\journal{}
\DeclareMathOperator{\Tr}{Tr}
\begin{document}
\begin{frontmatter}
\title{QCD $\theta$-vacuum in a Uniform Magnetic Field}
\author{Prabal Adhikari}
\ead{adhika1@stolaf.edu}
\address{Physics Department, Faculty of Natural Sciences and Mathematics, St. Olaf College, 1520 St. Olaf Avenue, Northfield, MN 55057, United States}
\date{\today}
\begin{abstract}
\noindent
We study the $\theta$-vacuum of QCD using two-flavor chiral perturbation theory ($\chi$PT) in the presence of a uniform, background magnetic field calculating the magnetic field-dependent free energy density, the topological density, the topological susceptibility and the fourth cumulant at one-loop order. We find that the topological susceptibility is enhanced by the magnetic field while the fourth topological cumulant is diminished at weak fields and enhanced at larger fields when $\theta=0$. However, in the QCD vacuum with $\theta\neq 0$, the topological susceptibility can be either monotonically enhanced or diminished relative to their $\theta$-vacuum values. The fourth cumulant also exhibits monotonic enhancement or suppression except for regions of $\theta$ near $0$ and $2\pi$, where it is both diminished and enhanced. Finally, the topological density is enhanced for all magnetic fields with its relative shift being identical to the relative shift of the up and down quark condensates in the $\theta$-vacuum.
\end{abstract}
\end{frontmatter}
\section{Introduction}
The vacuum of quantum chromodynamics (QCD) possesses topological properties as a consequence of the axial $U(1)$ anomaly~\cite{Crewther:1977ce,Witten:1979vv,Veneziano:1979ec,DiVecchia:1980yfw,SHIFMAN1980493}, which is closely connected to the yet unresolved strong CP problem, see Ref.~\cite{Cohen:2018cyj} for a recent discussion. The $n$-flavor QCD Lagrangian in the chiral limit is symmetric under $SU(n)_{V}\times SU(n)_{A}$, which is broken dynamically by the QCD vacuum to $SU(n)_{V}$ giving rise to $2n^{2}-1$ Goldstone bosons. Furthermore, the QCD Lagrangian is also symmetric under $U(1)_{V}$, which ensures the conservation of baryon current. Finally and of most relevance to this paper, the QCD Lagrangian in the chiral limit is symmetric under $U(1)_{A}$, which would imply that all hadrons possess opposite parity partners. However, the lightest quarks have small masses that break $SU(n)_{V}$ and give rise to pseudo-Goldstone bosons (also pseudo-scalars) instead of massless bosons. But there are no scalar counterparts, not even approximate ones, and the $U(1)_{A}$ symmetry is not spontaneously broken since this should give rise to a pseudoscalar isosinglet~\cite{Weinberg:1975ui}, suggesting that the $U(1)_{A}$ symmetry of the Lagrangian is broken strongly via a different mechanism, namely instantons~\cite{tHooft:1976rip, tHooft:1976snw}. Unlike the QCD Lagrangian (in the massless limit), the QCD partition function is not symmetric since the path integral measure transforms under $U(1)_{A}$ rotations~\cite{Fujikawa:1979ay}. As a consequence of this $U(1)_{A}$ anomaly, there exists a $\theta$-term in the QCD partition function
\begin{equation}
\begin{split}
Z=\int \mathcal{D}A\mathcal{D}q\mathcal{D}\bar{q}\exp\left [i\int d^{4}x\mathcal{L}_{\rm QCD} \right ]\ ,
\end{split}
\end{equation}
where the functional integral $\mathcal{D}q\mathcal{D}\bar{q}\equiv\prod_{f} \mathcal{D}q_{f}\bar{q}_{f}$, and the QCD Lagrangian including the $\theta$-term is
\begin{equation}
\begin{split}
\mathcal{L}_{\rm QCD}=&-\frac{1}{4}G_{\mu\nu}^{a}G_{a\mu\nu}-\frac{g^{2}\theta }{32\pi^{2}}\tilde{G}^{a\mu\nu}G_{\mu\nu}^{a}+\bar{q}\left (i\slashed{D}-M \right )q\ .
\end{split}
\end{equation}
$G_{\mu\nu}^{a}=\partial_{\mu}A_{\nu}^{a}-\partial_{\nu}A^{a}_{\mu}+gf_{abc}A^{b}_{\mu}A_{\nu}^{c}$ is the gluon field tensor, $\tilde{G}^{a}_{\mu\nu}=\frac{1}{2}\epsilon_{\mu\nu\alpha\beta} G^{a\alpha\beta}$ is the dual field strength tensor, the covariant derivative is $\slashed{D}_{\mu}=\slashed{\partial}_{\mu}-ig\slashed{A}^{a}_{\mu}\tfrac{\lambda^{a}}{2}$, $M$ is the diagonal, real quark mass matrix and $q$ is the quark field with flavor, Dirac and color indices suppressed. In the presence of an external magnetic field, the covariant derivative is modified to $\slashed{D}\rightarrow \slashed{D}-ieQ\slashed{A}$, where $Q$ is the diagonal quark charge matrix and $A^{\mu}$ is the external electromagnetic gauge field. Under an axial $U(1)$ rotation the quark fields transform as
\begin{equation}
\begin{split}
q\rightarrow e^{-i\Theta\gamma_{5}}q\ ,
\end{split}
\end{equation} 
with a corresponding transformation of the integration measure in the partition function~\cite{Fujikawa:1979ay} 
\begin{equation}
\begin{split}
\mathcal{D}\bar{q}\mathcal{D}q\rightarrow\exp\left [-i\int d^{4}x\frac{g^{2}\Theta n}{16\pi^{2}}\tilde{G}^{a\mu\nu}G_{a\mu\nu} \right ]\mathcal{D}\bar{q}\mathcal{D}q\ ,
\end{split}
\end{equation}
where the additional factor of $n$ arises due to the transformation of quark (and anti-quark) field measures associated with each flavor. The diagonal mass term in the Lagrangian, on the other hand, transforms as
\begin{equation}
\begin{split}
\mathcal{L}_{\rm mass}&=-\sum_{f=1}^{n}m_{f}\bar{q}_{f}q_{f}\rightarrow -\sum_{f=1}^{n}m_{f}\bar{q}_{f}e^{-i2\Theta\gamma_{5}}q_{f}\ ,
\end{split}
\end{equation}
where $m_{f}$ is the mass of the quark with flavor $f$.
Choosing $\Theta=-\frac{\theta}{2n}$ removes the explicit $\theta$ dependent term containing the dual field strength tensor in the Lagrangian at the expense of modifying the quark mass matrix term.~\footnote{For an elementary discussion of the structure of the QCD Lagrangian in the presence of a complex mass term, see for instance Ref.~\cite{srednicki2007quantum}} 

Since the contribution of the $\theta$-vacuum can be codified into the modified mass term in QCD and this rotation is unaffected by the presence of a $U(1)$ vector potential, we can use $\chi$PT with a modified scalar source to study the effects of the magnetic field on the topological cumulants. For quark masses and magnetic fields (strictly speaking $\sqrt{eH}$, where $H$ is the external field) that are small compared to the typical hadronic scale, $\Lambda_{\rm Had}\sim 4\pi f_{\pi}$, where $f_{\pi}$ is the pion decay constant, $\chi$PT~\cite{Gasser:1983yg,Gasser:1984gg,scherer2011primer} provides an effective field theoretic description with corrections that are systematically controlled. As such they have been previously used to characterize the topological susceptibility and other cumulants~\cite{Mao:2009sy,Guo:2015oxa,Bernard:2012ci,Leutwyler:1992yt}. The goal of this paper is to include the effect of a uniform background magnetic field. 

QCD has been studied in a background magnetic field due to their relevance to a wide range of phenomenological settings including magnetars, the quark-gluon plasma of the early universe and more recently for heavy-ion collisions. In the former two, the focus has been on the nature of chiral symmetry breaking while in the latter the focus has been on the chiral magnetic effect (CME), which leads to charge separation due to an external magnetic field and a chirality imbalance induced either by an electric field or an axial chemical potential (i.e. $\dot\theta$, the time evolution of the vacuum angle). While the properties of the QCD vacuum in the presence of an axial chemical potential (see for instance Ref.~\cite{Ruggieri:2020qtq}), has been studied, to the best of our knowledge, the effect of the magnetic field on the axial properties (as characterized by topological cumulants) in the confined phase of QCD appears to have been ignored. While the $\theta$-parameter of the universe is small as suggested by the experimental constraints on the neutron dipole moment~\cite{Afach:2015sja} (with a recent estimate of $\theta\lesssim10^{-11}$~\cite{kim2010axions}), we need not restrict ourselves to the $\theta=0$ vacuum -- in heavy ion collisions, $\theta$ is expected to be finite for small time scales within the QGP fireball~\cite{Buckley:1999mv}. Furthermore, one of the proposed solutions to the strong CP problem, i.e. the axion resolution~\cite{Weinberg:1977ma,Wilczek:1977pj}, involves a finite $\theta$ parameter relaxing to zero~\cite{Peccei:1977hh,Peccei:1977ur}. 

The paper is organized as follows: we begin with the calculation of the two-flavor free energy density at $\mathcal{O}(p^{4})$ in Section~\ref{sec:2f}, which we use to calculate the topological susceptibility, the fourth cumulant, the up and down quark condensates and susceptibilites. In Section~\ref{sec:discussion}, we characterize the relative shift of the topological susceptibility and fourth cumulant for general values of $\theta$. We also discuss low energy theorems that relate the shift in the topological susceptibility and fourth cumulant to that of the chiral condensate and susceptibility (for $\theta=0$) and the the relation of the relative shift in the topological susceptibility to that of the chiral condensate before concluding the paper with a summary, some final thoughts and speculations.

\section{Two-Flavor $\chi$PT}
\label{sec:2f}
\begin{figure}
	\centering
	\begin{subfigure}[b]{0.48\textwidth}
	\includegraphics[width=\textwidth]{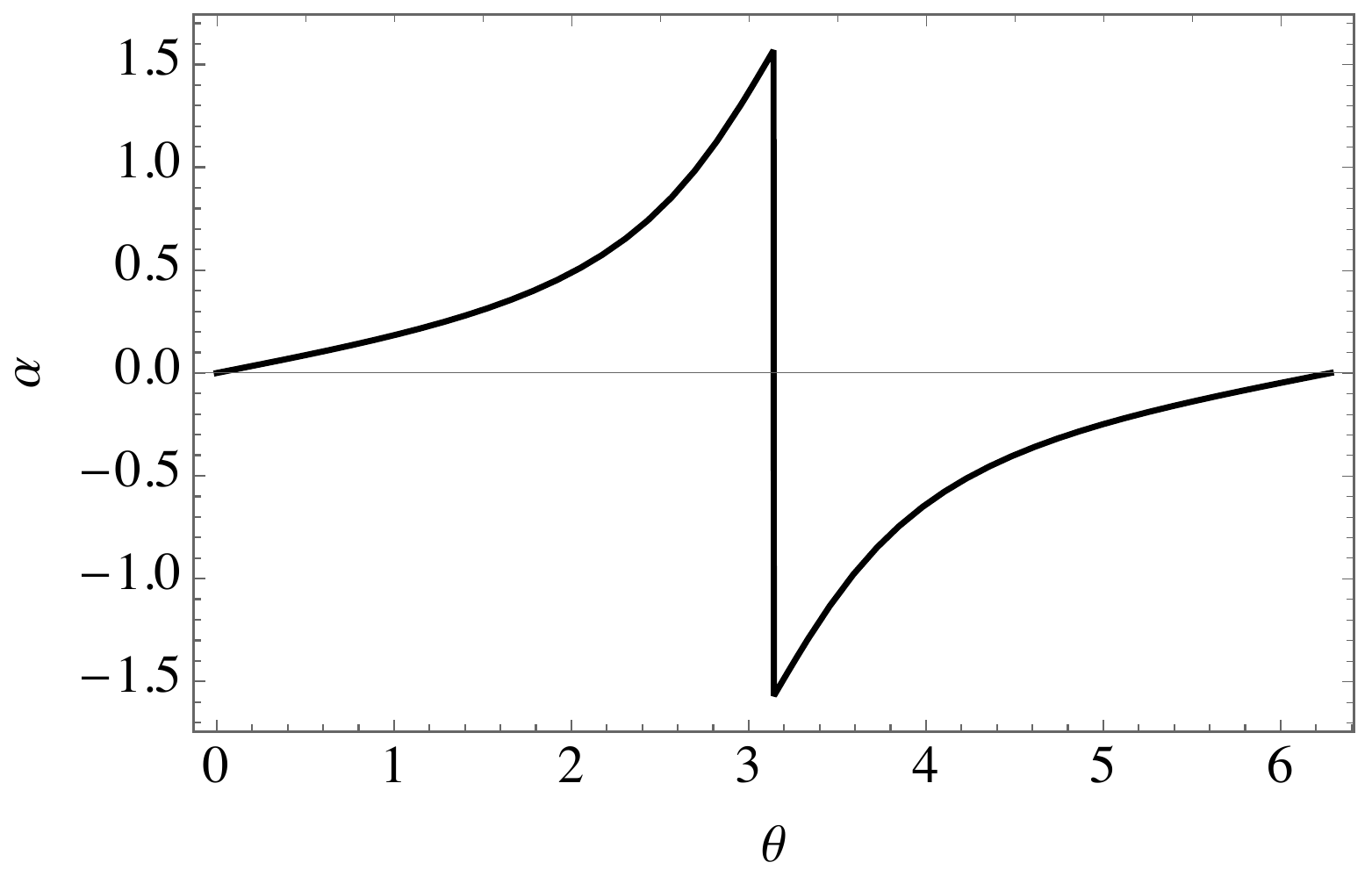}
	\end{subfigure}
	\begin{subfigure}[b]{0.48\textwidth}
	\includegraphics[width=\textwidth]{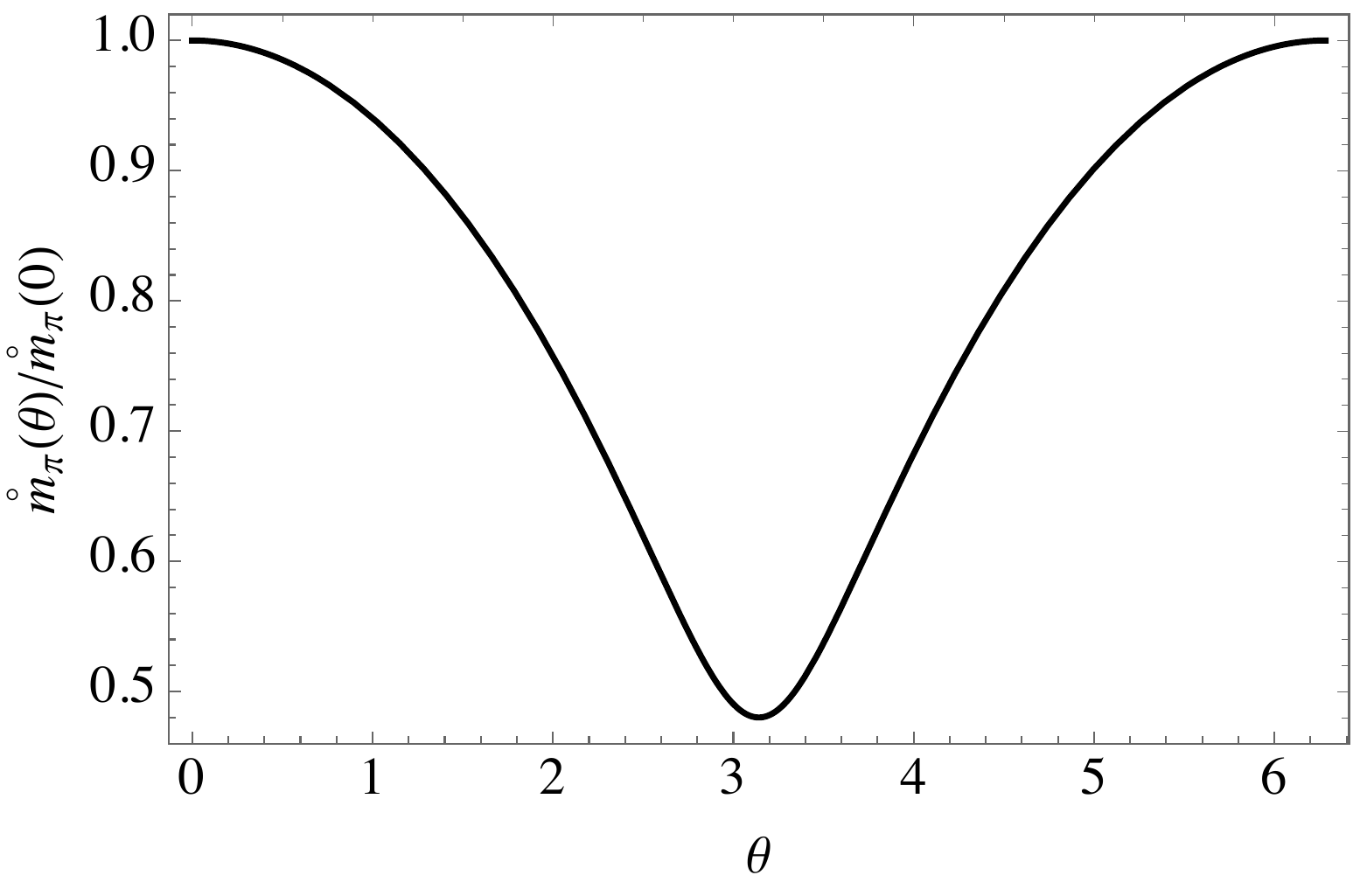}
	\end{subfigure}
\caption{Left: Plot of the ground state value of $\alpha$ for $0\le \theta\le 2\pi$. Right: Plot of the tree-level pion mass (normalized by the tree-level pion mass when $\theta=0$) for PDG values of quark masses given in Eq.~(\ref{eq:massPDG}).}
	\label{fig:gs-pionmass}
\end{figure}

\noindent
We begin our analysis with the $\mathcal{O}(p^{2})$ Lagrangian~\cite{Gasser:1983yg,scherer2011primer}
\begin{equation}
\begin{split}
\mathcal{L}_{2}&=-\frac{1}{4}F_{\mu\nu}F^{\mu\nu}+\frac{f^{2}}{4}\Tr\left [\nabla_{\mu}\Sigma(\nabla^{\mu}\Sigma)^{\dagger} \right ]+\frac{f^{2}}{4}\Tr\left[\chi\Sigma^{\dagger}+\Sigma\chi^{\dagger} \right ]\ ,
\end{split}
\end{equation}
where $\Sigma$ is an $SU(n)$ matrix, $f$ is the tree-level pion decay constant, $\chi$ is the scalar-pseudoscalar source and $F$ is the electromagnetic tensor. The covariant derivative is defined as
\begin{equation}
\begin{split}
\nabla_{\mu}\Sigma=\partial_{\mu}\Sigma-ieA_{\mu}[Q,\Sigma]\ ,
\end{split}
\end{equation} 
where $Q={\rm diag}\left(+\tfrac{2}{3},-\tfrac{1}{3}\right)$ is the quark charge matrix.
In the $\theta$-vacuum,
\begin{equation}
\begin{split}
\chi&=2Be^{-i\theta/n}M\ ,
\end{split}
\end{equation}
where $n$ is the number of quark flavors.
In the two-flavor case, the mass matrix,
\begin{equation}
\begin{split}
M&={\rm diag}(m_{u},m_{d})=\tfrac{1}{2}(m_{u}+m_{d})\mathbb{1}+\tfrac{1}{2}(m_{u}-m_{d})\tau_{3}\ .
\end{split}
\end{equation}
points in the $\mathbb{1}$ and $\tau_{3}$ directions. As such, we anticipate the possibility of the ground state, $\Sigma_{\alpha}$, also pointing in the $\tau_{3}$ direction. In order to study this explicity, we proceed by parameterizing the most general form for the ground state in the presence of the $\theta$ term
\begin{equation}
\begin{split}
\Sigma_{\alpha}=\cos\alpha\ \mathbb{1}+i\sin\alpha\ \hat{\phi}_{i}\tau_{i}\ ,
\end{split}
\end{equation}
where we adopt the Einstein summation convention with an implied sum over the isospin index $a$ and $\hat{\phi}_{a}\hat{\phi}_{a}=1$ which guarantees unitarity, i.e. $\Sigma_{\alpha}^{\dagger}\Sigma_{\alpha}=\mathbb{1}$. Then the tree-level free energy (excluding the external magnetic field) is
\begin{equation}
\begin{split}
\tilde{\mathcal{F}}_{\rm tree}=&-f^{2}B\left[(m_{u}+m_{d})\cos\alpha\cos\tfrac{\theta}{2}+(m_{d}-m_{u})\hat{\phi}_{3}\sin\alpha\sin\tfrac{\theta}{2}\right]\ ,
\end{split}
\end{equation}
which is minimized when $\hat{\phi}_{3}=1$, assuming $m_{d}>m_{u}$. Then $\hat{\phi}_{1}=\hat{\phi}_{2}=0$ and
\begin{equation}
\begin{split}
\label{eq:Ftreet}
\tilde{\mathcal{F}}_{\rm tree}&=-f^{2}B\left[m_{u}\cos\phi_{u}(\theta)+m_{d}\cos\phi_{d}(\theta)\right]\ ,
\end{split}
\end{equation}
where $\phi_{u}(\theta)=\tfrac{\theta}{2}+\alpha(\theta)$ and $\phi_{d}(\theta)=\tfrac{\theta}{2}-\alpha(\theta)$ with their values adding to the vacuum angle, $\theta$. $\alpha$ is found by minimizing the tree-level free energy, $\tilde{\mathcal{F}}_{\rm tree}$,
\begin{equation}
\begin{split}
\label{eq:alphags}
\tan\alpha&=\tfrac{m_{d}-m_{u}}{m_{u}+m_{d}}\tan\tfrac{\theta}{2}\ ,
\end{split}
\end{equation}
which is a well-known result in $\chi$PT, for instance see Ref.~\cite{srednicki2007quantum}. In the left panel of Fig.~\ref{fig:gs-pionmass}, we plot the $\alpha$ for values of $\theta$ between $0$ and $2\pi$ assuming realistic values of quark masses with $m_{d}>m_{u}$ extracted from the Particle Data Group~\cite{Zyla:2020zbs}, see Eq.~(\ref{eq:massPDG}). The ground state value of $\alpha$ is non-zero for all values of $\theta$ except $0$, $\pi$ and $2\pi$. For $0<\theta<\pi$, $\alpha>0$ and for $\pi<\theta<2\pi$, $\alpha<0$.

In order to study the effect on the topological susceptibility due to a uniform magnetic field, which first appears at next-to-leading order in the chiral expansion, we proceed by parameterizing the fluctuations of the pion fields using 
\begin{equation}
\begin{split}
\Sigma&=\mathcal{A}_{\alpha}e^{i\frac{\phi_{a}\tau_{a}}{f}}\mathcal{A}_{\alpha}\textrm{, where }\mathcal{A}_{\alpha}=e^{i\frac{\alpha}{2}\tau_{3}}\\
\end{split}
\end{equation}
and there is an implied sum over the isospin index $a=1,2,3$ and $\phi_{a}$ are the fluctuations around the ground state value of $\alpha$, which is altered from the tree-level value of Eq.~(\ref{eq:alphags}) due to quantum fluctuations. Using the standard definition of charge eigenstates,
\begin{equation}
\begin{split}
\phi_{a}\tau_{a}&=
\begin{pmatrix}
\pi^{0}&\sqrt{2}\pi^{+}\\
\sqrt{2}\pi^{-}&-\pi^{0}\\
\end{pmatrix}\ ,
\end{split}
\end{equation}
the linear contribution to the Lagrangian is
\begin{align}
\label{eq:L2lin}
\mathcal{L}_{2,\rm linear}&=Bf[m_{u}\sin\phi_{u}(\theta)-m_{d}\sin\phi_{d}(\theta)]\pi^{0}
\end{align}
and the quadratic contribution is
\begin{align}
\label{eq:L2quad}
\mathcal{L}_{2,\rm quad}&=-\frac{1}{2}H^{2}+D_{\mu}\pi^{+}D^{\mu}\pi^{-}-\mathring{m}_{\pi^{\pm}}^{2}(\theta)\pi^{+}\pi^{-}+\frac{1}{2}\partial_{\mu}\pi^{0}\partial^{\mu}\pi^{0}-\frac{1}{2}\mathring{m}_{\pi^{0}}^{2}(\theta)\pi^{0}\pi^{0}\ ,
\end{align}
where $D_{\mu}\pi^{\pm}\equiv(\partial_{\mu}\pm ieA_{\mu})\pi^{\pm}$ and the tree-level charged and neutral pion masses, $\mathring{m}_{\pi^{\pm}}^{2}(\theta)$ and $\mathring{m}_{\pi^{0}}^{2}(\theta)$, respectively are degenerate,
\begin{align}
\label{eq:pionmass}
\mathring{m}_{\pi^{\pm}}^{2}(\theta)&=\mathring{m}_{\pi^{0}}^{2}(\theta)=B\left [ m_{u}\cos\phi_{u}(\theta)+m_{d}\cos\phi_{d}(\theta) \right ]=B\sqrt{m_{u}^{2}+m_{d}^{2}+2m_{u}m_{d}\cos\theta}\ ,
\end{align}
with the second equality following from using the tree-level value of $\alpha(\theta)$ from Eq.~(\ref{eq:alphags}). In the right panel of Fig.~\ref{fig:gs-pionmass}, we plot the ratio of the degenerate pion mass, $\mathring{m}_{\pi}(\theta)$ in the $\theta$ vacuum to the pion mass in the $\theta=0$ vacuum. The pion mass is even about $\theta=0$ with a minimum at $\theta=\pi$. This is straightforward to note by factoring out $(m_{u}+m_{d})^{2}$ from the square root in Eq.~(\ref{eq:pionmass}) and identifying $\mathring{m}_{\pi}(0)$ with $\sqrt{B(m_{u}+m_{d})}$. The resulting expression
\begin{align}
\label{eq:pionmass2}
\mathring{m}_{\pi}(\theta)=\mathring{m}_{\pi}(0)\sqrt[4]{1-\tfrac{4m_{u}m_{d}}{(m_{u}+m_{d})^{2}}\sin^{2}\tfrac{\theta}{2}}\ ,
\end{align}
makes transparent the salient features depicted in the figure.

For a next-to-leading order calculation that incorporates the effects of a background magnetic field, we need the one loop contributions arising from $\mathcal{L}_{2,\rm quad}$, which are of $\mathcal{O}(p^{4})$ and the tree-level contributions that arise from $\mathcal{L}_{4}$. The divergences from the latter exactly cancel the one-loop divergences. For two-flavor $\chi$PT, we need the tree-level contribution from the $\mathcal{O}(p^{4})$ Lagrangian
\begin{equation}
\begin{split}
\mathcal{L}_{4}&=\frac{l_{3}+l_{4}}{16}\Tr\left[\chi\Sigma^{\dagger}+\chi^{\dagger}\Sigma\right]^{2}+\frac{h_{1}+h_{3}-l_{4}}{4}\Tr(\chi \chi^{\dagger})-\frac{l_{7}}{16}\left [\Tr(\chi\Sigma^{\dagger}-\chi^{\dagger}\Sigma)\right ]^{2}\\
&+\frac{h_{1}-h_{3}-l_{4}}{16}\left\{\left [\Tr(\chi\Sigma^{\dagger}+\chi^{\dagger}\Sigma)\right ]^{2}+\left [\Tr(\chi\Sigma^{\dagger}-\chi^{\dagger}\Sigma)\right ]^{2}-2\Tr \left(\Sigma \chi^{\dagger}\Sigma \chi^{\dagger}+\chi\Sigma^{\dagger}\chi\Sigma^{\dagger} \right )\right\}\\
&-\frac{4h_{2}+l_{5}}{2}\Tr\left [F_{\mu\nu}^{R}F^{R\mu\nu}+F_{\mu\nu}^{L}F^{L\mu\nu} \right ]+l_{5}\Tr\left [\Sigma F_{\mu\nu}^{L}\Sigma^{\dagger}F^{R\mu\nu}\right ]\ ,
\end{split}
\end{equation}
where $F_{\mu\nu}^{R}=F^{L}_{\mu\nu}=-e\frac{\tau_{3}}{2}F_{\mu\nu}$ and $F_{\mu\nu}$ is the electromagnetic tensor.
The low energy constants, on the other hand, required for renormalization are
\begin{align}
\label{LEC}
l_{i}&=l_{i}^{r}+\gamma_{i}\lambda,\ h_{i}=h_{i}^{r}+\delta_{i}\lambda,\ \lambda=-\frac{\Lambda^{-2\epsilon}}{2(4\pi)^{2}}\left(\frac{1}{\epsilon}+1\right)\ ,
\end{align}
where both $l_{i}^{r}$ and $h_{i}^{r}$ are scale-dependent. They are defined as
\begin{equation}
\begin{split}
l_{i}^{r}&=\frac{\gamma_{i}}{2(4\pi)^{2}}\left [\bar{l}_{i}+\log\frac{2B\hat{m}}{\Lambda^{2}} \right ],\ h_{i}^{r}=\frac{\delta_{i}}{2(4\pi)^{2}}\left [\bar{h}_{i}+\log\frac{2B\hat{m}}{\Lambda^{2}} \right ]\ ,
\end{split}
\end{equation}
where $\hat{m}=\frac{m_{u}+m_{d}}{2}$ is the average light quark mass, $\Lambda$ is the renormalization scale in the $\overline{\rm MS}$-scheme and the constants $\gamma_{i}$ and $\delta_{i}$ required for renormalization are
\begin{align}
\gamma_{3}=-\frac{1}{2},\ \gamma_{4}&=2,\ \gamma_{5}=-\frac{1}{6},\ \gamma_{7}=0,\ \delta_{1}=2,\ \delta_{2}=\frac{1}{12},\ \delta_{3}=0\ ,
\end{align}
with the definitions of $l_{i}^{r}$ and $h_{i}^{r}$ suspended when either $\gamma_{i}$ or $\delta_{i}$ are zero.

With all the ingredients for renormalization in place, we next require the one-loop contribution to the effective potential. For the charged pion, this requires summing over the Landau energy levels. The contribution to the effective potential for a pair of charged meson with bare mass $m$ is
\begin{equation}
\begin{split}
I_{H}(m)&=\frac{eH}{2\pi}\sum_{k=0}^{\infty}\int_{p_{0},p_{z}}\ln[p_{0}^{2}+p_{z}^{2}+m_{H}^{2}]\\
\end{split}
\end{equation}
where $m_{H}^{2}=m^{2}+(2k+1)|eH|$ and $\int_{p_{0}p_{z}}\equiv\int\frac{dp_{0}}{2\pi}\frac{dp_{z}}{2\pi}$. We proceed by first taking the derivative with respect to $m^{2}$, introducing Schwinger's proper time variable, $s$, summing over all the Landau level, i.e. $k$, and finally integrating over $p_{0}$, $p_{z}$ and $m^{2}$. We get
\begin{equation}
\begin{split}
I_{H}(m)&=-\frac{\mu^{2\epsilon}}{(4\pi)^{2}}\int_{0}^{\infty} ds\ \frac{1}{s^{3-\epsilon}}e^{-m^{2}s}\left[\frac{eHs}{\sinh eHs}\right]\ ,
\end{split}
\end{equation}
where we have written in the Schwinger proper-time integral in $d=2-2\epsilon$ dimensions with $\mu=\sqrt{e^{\gamma_{E}}\Lambda^{2}}$, which is convenient for dimensional regularization. 
We note that the integral is divergent in the ultraviolet: for small values of $s$, the integrand diverges as $s^{-3}$ even when the external magnetic field is absent. Isolating both the $m$ and $H$ dependent divergences we get,
\begin{align}
I_{H}(m)=&I_{H}^{\rm div}(m)+I_{H}^{\rm fin}(m)\\
I_{H}^{\rm div}(m)=&-\frac{m^{4}}{2(4\pi)^{2}}\left[\frac{1}{\epsilon}+\frac{3}{2}+\log\frac{\Lambda^{2}}{m^{2}} \right]+\frac{(eH)^{2}}{6(4\pi)^{2}}\left[\frac{1}{\epsilon}+\log\frac{\Lambda^{2}}{m^{2}}\right]\\
\label{eq:IHfin}
I_{H}^{\rm fin}(m)=&-\frac{1}{(4\pi)^{2}}\int_{0}^{\infty} \frac{ds}{s^{3}}e^{-m^{2}s}\left[\frac{eHs}{\sinh eHs}-1+\frac{(eHs)^{2}}{6}\right]\ .
\end{align}
We have not only isolated the standard $m^{4}$ divergence in the absence of an external magnetic field but the divergent $H$-dependent contribution ensures that the finite term, $I_{H}^{\rm fin}$, is consistent with that of Ref.~\cite{Schwinger:1951nm} -- it also turns out to be the most convenient choice for charge renormalization.

We now calculate the 1-loop free energy at finite $H$ in terms of $\phi_{u}(\theta)$ and $\phi_{d}(\theta)$ using the tree-level contribution to the free energy $\mathcal{F}_{\rm tree}$, the contribution from one-loop graphs, $\mathcal{F}_{1}$ and the counterterms, $\mathcal{F}_{\rm ct}$,
\begin{align}
\mathcal{F}_{\rm tree}(\theta)&=\frac{1}{2}H^{2}-f^{2}B\left[m_{u}\cos\phi_{u}(\theta)+m_{d}\cos\phi_{d}(\theta)\right]\\
\mathcal{F}_{1}(\theta)&=I_{H}[\mathring{m}_{\pi_{\pm}}(\theta)]+\frac{1}{2}I_{0}[\mathring{m}_{\pi^{0}}(\theta)]\\\nonumber
\mathcal{F}_{\rm ct}(\theta)&=-(l_{3}+l_{4})\left [B\left\{m_{u}\cos\phi_{u}(\theta)+m_{d}\cos\phi_{d}(\theta)\right\} \right ]^{2}\\\nonumber
&-l_{7}\left [B\left\{-m_{u}\sin\phi_{u}(\theta)-m_{d}\sin\phi_{d}(\theta)\right\}\right ]^{2}-(h_{1}+h_{3}-l_{4})\left[B^{2}(m_{u}^{2}+m_{d}^{2})\right]\\
&-2(h_{1}-h_{3}-l_{4})\left[B^{2}m_{u}m_{d}\cos\theta\right]+4h_{2}(eH)^{2}\ ,
\end{align}
where $\mathcal{F}_{\rm tree}=\frac{1}{2}H^{2}+\tilde{\mathcal{F}}_{\rm tree}$, is the full tree-level contribution to the free energy including the external magnetic field, with $\tilde{\mathcal{F}}_{\rm tree}$ is the contribution due to the $H=0$ vacuum from Eq.~(\ref{eq:Ftreet}). $\mathcal{F}_{1}$ is the contribution to the free energy arising through one-loop diagrams -- the first term arises through the interaction of the charged pions with the external magnetic field and the second through the neutral pion, which does not interact with the magnetic field. Finally, $\mathcal{F}_{\rm ct}$ represents the tree-level counterterms and has divergences that arise through the low-energy-constants of Eq.~(\ref{LEC}). These are cancelled exactly by the ones arising through the one-loop contributions, which includes divergences quadratic in the magnetic field. The resulting one-loop free energy in the presence of the $\theta$-term and a uniform magnetic field can be written in terms of $H$-independent and dependent contributions
\begin{equation}
\begin{split}
\mathcal{F}(\theta,H)=&\mathcal{F}_{0}(\theta)+\mathcal{F}_{H}(\theta)\ .
\end{split}
\end{equation}
where the $H$-independent contribution is
\begin{align}
\label{eq:F0-2f}
\nonumber
&\mathcal{F}_{0}(\theta)=-f^{2}\mathring{m}_{\pi}^{2}(\theta)-(l^{r}_{3}+l^{r}_{4})\mathring{m}_{\pi}^{4}(\theta)-\frac{3\mathring{m}_{\pi}^{4}(\theta)}{4(4\pi)^{2}}\left [\frac{1}{2}+\log\frac{\Lambda^{2}}{\mathring{m}_{\pi}^{2}(\theta)} \right]\\
\nonumber
&-(h^{r}_{1}-l^{r}_{4}-h_{3})\left[B^{2}\{m_{u}^{2}+m_{d}^{2}+2m_{u}m_{d}\cos\theta\}\right]\\
&-2h_{3}\left [B^{2}\{m_{u}^{2}+m_{d}^{2}\}\right ]-l_{7}\left [B\left\{m_{u}\sin\phi_{u}(\theta)+m_{d}\sin\phi_{d}(\theta)\right\}\right ]^{2}\ ,
\end{align}
which is in agreement with Ref.~\cite{Guo:2015oxa}. The magnetic field dependent contribution is
\begin{equation}
\begin{split}
\mathcal{F}_{H}(\theta)&=\frac{1}{2}H_{R}^{2}+\frac{(eH)^{2}}{(4\pi)^{2}}\mathfrak{I}_{H}^{\rm}(\tfrac{\mathring{m}^{2}_{\pi}(\theta)}{eH})\ ,
\end{split}
\end{equation}
where $H_{R}=Z_{H}H$ is the renormalized magnetic field, $\mathfrak{I}_{H}$ is the Schwinger integral defined in Eq.~(\ref{Schwingerint}) and $Z_{H}^{-1}$ is the charge renormalization wave function, which ensures that $eH$ remains invariant,
\begin{equation}
\begin{split}
\label{cumulant-def}
Z_{H}&=1+4e^{2}h^{r}_{2}+\frac{e^{2}}{6(4\pi)^{2}}\left(\log\frac{\Lambda^{2}}{\mathring{m}_{\pi}^{2}(\theta)}-1\right)\ .
\end{split}
\end{equation}

The $\theta$-dependence in $\mathcal{F}_{0}$ enters through $\phi_{u}(\theta)=\tfrac{\theta}{2}+\alpha$ and $\phi_{d}(\theta)=\tfrac{\theta}{2}-\alpha$, where $\alpha$ is the vacuum orientation that also depends on $\theta$ and is modified from its tree-level value of Eq.~(\ref{eq:alphags}) due to one-loop effects. This change in the value of $\alpha$ is straightforward to calculate at one loop level for all values of $\theta$ and depends only on the LEC $l_{7}$ since all other terms in the NLO free energy depend on $\theta$ only through $\mathring{m}_{\pi}(\theta)$:  the leading value of $\alpha$ is found by setting the derivative of $-f^{2}\mathring{m}^{2}_{\pi}(\theta)$ to zero. In the next-to-leading order calculation of $\alpha$, this ensures that terms that depend on $\theta$ purely through $\mathring{m}_{\pi}(\theta)$ do not contribute to an NLO change in the tree-level ground state value of $\alpha$. Since all contributions in the free energy including terms-dependent on $H$ do, with the exception of the term proportional to $l_{7}$,  the NLO value of $\alpha$ depends only on $l_{7}$~\cite{Acharya:2015pya}. However, this change in the value of $\alpha$ does not contribute to the free energy at $\mathcal{O}(p^{4})$. We, therefore get, including the $H$-dependent contribution, $\mathcal{F}_{H}$, the following free energy
\begin{equation}
\begin{split}
\mathcal{F}(\theta,H)=&\frac{1}{2}H_{R}^{2}-f^{2}\mathring{m}^{2}_{\pi}(\theta)-(l^{r}_{3}+h^{r}_{1}-h_{3})\mathring{m}_{\pi}^{4}(\theta)-\frac{3\mathring{m}_{\pi}^{4}(\theta)}{4(4\pi)^{2}}\left [\frac{1}{2}+\log\frac{\Lambda^{2}}{\mathring{m}_{\pi}^{2}(\theta)} \right ]\\
&-2h_{3}\left [B^{2}\{m_{u}^{2}+m_{d}^{2}\}\right ]-l_{7}\frac{4B^{4}m_{u}^{2}m_{d}^{2}\sin^{2}\theta}{\mathring{m}_{\pi}^{4}(\theta)}+\frac{(eH)^{2}}{(4\pi)^{2}}\mathfrak{I}_{H}(\tfrac{\mathring{m}_{\pi}^{2}(\theta)}{eH})\ ,
\end{split}
\end{equation}
where $\mathring{m}_{\pi}(\theta)$ is the same as in Eq.~(\ref{eq:pionmass2}). The term proportional to $l_{7}$ contributes to the free energy at finite $\theta$ mod $\pi$.

\subsection{Topological Density, Susceptibility and Fourth Cumulant}
\noindent
The free energy can be used to calculate the topological density, susceptibility and fourth cumulant up to $\mathcal{O}(p^{4})$. They are defined in terms of the full free energy, $\mathcal{F}$ as
\begin{equation}
\begin{split}
\rho_{t}=\frac{\partial\mathcal{F}(\theta,H)}{\partial\theta}\ ,\ \chi_{t}=\frac{\partial^{2}\mathcal{F}(\theta,H)}{\partial \theta^{2}}\ ,\ c_{4}=\frac{\partial^{4}\mathcal{F}(\theta,H)}{\partial \theta^{4}}\ ,
\end{split}
\end{equation}
respectively.  Using the definitions, we get for the topological density
\begin{equation}
\begin{split}
\label{eq:topden}
&\rho_{t}(\theta,H)=\frac{B^{2}f^{2}m_{u}m_{d}\sin\theta}{\mathring{m}_{\pi}^{2}(\theta)}+2B^{2}m_{u}m_{d}\sin\theta(l^{r}_{3}+h^{r}_{1}-h_{3})+\frac{3B^{2}m_{u}m_{d}}{2(4\pi)^{2}}\log\frac{\Lambda^{2}}{\mathring{m}_{\pi}^{2}(\theta)}\sin\theta\\
-&8l_{7}\left(\frac{B^{2}m_{u}m_{d}}{\mathring{m}_{\pi}^{2}(\theta)}\right)^{2}\sin\theta\left[\cos\theta+\frac{B^{2}m_{u}m_{d}}{\mathring{m}_{\pi}^{4}(\theta)}\sin^{2}\theta\right]-\frac{(eH)^{2}}{(4\pi)^{2}}\frac{B^{2}m_{u}m_{d}}{\mathring{m}_{\pi}^{2}(\theta)}\sin\theta\mathcal{I}_{H,2}(\tfrac{\mathring{m}_{\pi}^{2}(\theta)}{eH})\ ,
\end{split}
\end{equation}
with each term proportional to $\sin\theta$ and suggesting that the topological density vanishes when $\theta=0$ and becomes finite for $\theta>0$ and excluding $\theta=\pi$ and $2\pi$ with the latter equivalent to $\theta=0$. The topological susceptibility, on the other hand, is
\begin{equation}
\begin{split}
\label{eq:chit}
&\chi_{t}(\theta,H)=\frac{B^{2}f^{2}m_{u}m_{d}}{\mathring{m}_{\pi}^{2}(\theta)}\left(\cos\theta+\frac{B^{2}m_{u}m_{d}}{\mathring{m}_{\pi}^{4}(\theta)}\sin^{2}\theta\right)+2B^{2}(l^{r}_{3}+h^{r}_{1}-h_{3})m_{u}m_{d}\cos\theta\\
&-\frac{3B^{2}m_{u}m_{d}}{2(4\pi)^{2}}\left[\frac{B^{2}m_{u}m_{d}\sin^{2}\theta}{\mathring{m}_{\pi}^{4}(\theta)}+\cos\theta\log\frac{\Lambda^{2}}{\mathring{m}_{\pi}^{2}(\theta)}\right]\\
&-l_{7}\frac{8B^{4}m_{u}^{2}m_{d}^{2}}{\mathring{m}_{\pi}^{4}(\theta)}\left[\cos2\theta+\frac{5B^{2}m_{u}m_{d}}{\mathring{m}_{\pi}^{4}(\theta)}\cos\theta\sin^{2}\theta+\frac{4B^{4}m_{u}^{2}m_{d}^{2}}{\mathring{m}_{\pi}^{8}(\theta)}\sin^{4}\theta\right ]\\
&-\frac{eH}{(4\pi)^{2}}\frac{B^{2}m_{u}m_{d}}{\mathring{m}_{\pi}^{2}(\theta)}\left(\cos\theta+\frac{B^{2}m_{u}m_{d}}{\mathring{m}_{\pi}^{4}(\theta)}\sin^{2}\theta\right)\mathcal{I}_{H,2}(\tfrac{\mathring{m}_{\pi}^{2}(\theta)}{eH})-\frac{1}{(4\pi)^{2}}\left(\frac{B^{2}m_{u}m_{d}}{\mathring{m}_{\pi}^{2}(\theta)}\sin^{2}\theta\right)\mathcal{I}_{H,1}(\tfrac{\mathring{m}_{\pi}^{2}(\theta)}{eH})\ ,
\end{split}
\end{equation}
and the fourth cumulant is
\begin{equation}
\begin{split}
\label{eq:c4}
c_{4}(\theta,H)&=-\frac{B^{2}f^{2}m_{u}m_{d}}{\mathring{m}_{\pi}^{2}(\theta)}\cos\theta\left(1-\frac{3B^{2}m_{u}m_{d}}{\mathring{m}_{\pi}^{4}(\theta)}\cos\theta\right)-2B^{2}(l^{r}_{3}+h^{r}_{1}-h_{3})m_{u}m_{d}\cos\theta\\
&-\frac{3B^{2}m_{u}m_{d}}{2(4\pi)^{2}}\left[\cos\theta\log\frac{\Lambda^{2}}{\mathring{m}_{\pi}^{2}(\theta)}+\frac{B^{2}m_{u}m_{d}}{\mathring{m}_{\pi}^{4}(\theta)}\left(-3\cos^{2}\theta+4\sin^{2}\theta\right)\right.\\
&\left.-\frac{4B^{4}m_{u}^{2}m_{d}^{2}\sin^{2}\theta}{\mathring{m}_{\pi}^{8}(\theta)}\left(3\cos\theta+\frac{2B^{2}m_{u}m_{d}}{\mathring{m}_{\pi}^{4}(\theta)}\sin^{2}\theta\right)\right ]\\
&+l_{7}\frac{32B^{4}m_{u}^{2}m_{d}^{2}}{\mathring{m}_{\pi}^{4}(\theta)}\left[\cos2\theta+\frac{B^{2}m_{u}m_{d}}{\mathring{m}_{\pi}^{4}(\theta)}\left\{-3\cos^{3}\theta+\frac{53}{4}\sin^{2}\theta\cos\theta\right\}\right.\\
&+\left.\left(\frac{B^{2}m_{u}m_{d}}{\mathring{m}_{\pi}^{4}(\theta)}\right)^{2}\left\{16\sin^{4}\theta-\frac{39}{4}\sin^{2}\theta\right\}\right.\\
&\left.-\left(\frac{B^{2}m_{u}m_{d}}{\mathring{m}_{\pi}^{4}(\theta)}\right)^{3}84\sin^{4}\theta\cos\theta-\left(\frac{B^{2}m_{u}m_{d}}{\mathring{m}_{\pi}^{4}(\theta)}\right)^{4}48\sin^{6}\theta\right]\\
&+\left(\frac{B^{2}m_{u}m_{d}}{\mathring{m}_{\pi}^{2}(\theta)}\right)\left[\cos\theta+\frac{B^{2}m_{u}m_{d}}{\mathring{m}_{\pi}^{4}(\theta)}(4\sin^{2}\theta-3\cos^{2}\theta)-\left(\frac{B^{2}m_{u}m_{d}}{\mathring{m}_{\pi}^{4}(\theta)}\right)^{2}9\sin\theta\sin2\theta\right.\\
&\left.-\left(\frac{B^{2}m_{u}m_{d}}{\mathring{m}_{\pi}^{4}(\theta)}\right)^{3}15\sin^{4}\theta\right]\frac{eH}{(4\pi)^{2}}\mathcal{I}_{H,2}(\tfrac{\mathring{m}^{2}_{\pi}(\theta)}{eH})\\
&+\left(\frac{B^{2}m_{u}m_{d}}{\mathring{m}_{\pi}^{2}(\theta)}\right)^{2}\left[(4\sin^{2}\theta-3\cos^{2}\theta)-\left(\frac{B^{2}m_{u}m_{d}}{\mathring{m}_{\pi}^{4}(\theta)}\right)9\sin\theta\sin2\theta\right.\\
&\left.-\left(\frac{B^{2}m_{u}m_{d}}{\mathring{m}_{\pi}^{4}(\theta)}\right)^{2}15\sin^{4}\theta\right ]\frac{1}{(4\pi)^{2}}\mathcal{I}_{H,1}(\tfrac{\mathring{m}_{\pi}^{2}(\theta)}{eH})\\
&-\left(\frac{B^{2}m_{u}m_{d}}{\mathring{m}_{\pi}^{2}(\theta)}\right)^{3}\left[6\sin^{2}\theta\left\{\cos\theta+\left(\frac{B^{2}m_{u}m_{d}}{\mathring{m}_{\pi}^{4}(\theta)}\right)\sin^{2}\theta\right\} \right ]\frac{1}{(4\pi)^{2}eH}\mathcal{I}_{H,0}(\tfrac{\mathring{m}_{\pi}^{2}(\theta)}{eH})\\
&-\left(\frac{B^{2}m_{u}m_{d}}{\mathring{m}_{\pi}^{2}(\theta)}\right)^{4}\sin^{4}\theta\frac{1}{(4\pi)^{2}(eH)^{2}}\mathcal{I}_{H,-1}(\tfrac{\mathring{m}_{\pi}^{2}(\theta)}{eH})\ ,
\end{split}
\end{equation}
where the integrals $\mathcal{I}_{H,2}(y)$, $\mathcal{I}_{H,1}(y)$, $\mathcal{I}_{H,0}(y)$ and $\mathcal{I}_{H,-1}(y)$ are defined in Eqs.~(\ref{eq:IH2}), (\ref{eq:IH1}), (\ref{eq:IH0}) and (\ref{eq:IH-1}) respectively. The pion mass $\mathring{m}_{\pi}(\theta)$ is defined in Eqs.~(\ref{eq:pionmass}) and (\ref{eq:pionmass2}).
\subsection{Quark Condensates and Susceptibilities}
\noindent
The topological observables can be related to the light quark condensates and susceptibilities. They are defined as 
\begin{equation}
\begin{split}
\langle\bar{q}_{f}q_{f}\rangle=\frac{\partial \mathcal{F}(\theta,H)}{\partial m_{q_{f}}}\ ,\ \chi_{q_{f}}=\frac{\partial^{2} \mathcal{F}(\theta,H)}{\partial m_{q_{f}}^{2}}
\end{split}
\end{equation}
where $q_{f}=u$ or $d$. The up and down quark condensates in the $\theta$-vacuum with a background magnetic field is
\begin{equation}
\begin{split}
\label{eq:uu}
\langle\bar{u}u\rangle(\theta,H)&=-\frac{B^{2}f^{2}(m_{u}+m_{d}\cos\theta)}{\mathring{m}_{\pi}^{2}(\theta)}-2B^{2}(m_{u}+m_{d}\cos\theta)(l^{r}_{3}+h^{r}_{1})-2B^{2}(m_{u}-m_{d}\cos\theta)h_{3}\\
&-\frac{3B^{2}(m_{u}+m_{d}\cos\theta)}{2(4\pi)^{2}}\log\frac{\Lambda^{2}}{\mathring{m}^{2}_{\pi}(\theta)}-l_{7}\frac{8B^{4}m_{u}m_{d}^{2}\sin^{2}\theta}{\mathring{m}_{\pi}^{4}(\theta)}\left(1-\frac{B^{2}m_{u}(m_{u}+m_{d}\cos\theta)}{\mathring{m}_{\pi}^{4}(\theta)}\right)\\
&+\frac{eH}{(4\pi)^{2}}\frac{B^{2}(m_{u}+m_{d}\cos\theta)}{\mathring{m}_{\pi}^{2}(\theta)}\mathcal{I}_{H,2}(\tfrac{\mathring{m}_{\pi}^{2}(\theta)}{eH})\ ,
\end{split}
\end{equation}
\begin{equation}
\begin{split}
\label{eq:dd}
\langle\bar{d}d\rangle(\theta,H)&=-\frac{B^{2}f^{2}(m_{d}+m_{u}\cos\theta)}{\mathring{m}_{\pi}^{2}(\theta)}-2B^{2}(m_{d}+m_{u}\cos\theta)(l^{r}_{3}+h^{r}_{1})-2B^{2}(m_{d}-m_{u}\cos\theta)h_{3}\\
&-\frac{3B^{2}(m_{d}+m_{u}\cos\theta)}{2(4\pi)^{2}}\log\frac{\Lambda^{2}}{\mathring{m}^{2}_{\pi}(\theta)}-l_{7}\frac{8B^{4}m_{d}m_{u}^{2}\sin^{2}\theta}{\mathring{m}_{\pi}^{4}(\theta)}\left(1-\frac{B^{2}m_{d}(m_{d}+m_{u}\cos\theta)}{\mathring{m}_{\pi}^{4}(\theta)}\right)\\
&+\frac{eH}{(4\pi)^{2}}\frac{B^{2}(m_{d}+m_{u}\cos\theta)}{\mathring{m}_{\pi}^{2}(\theta)}\mathcal{I}_{H,2}(\tfrac{\mathring{m}_{\pi}^{2}(\theta)}{eH})
\ ,
\end{split}
\end{equation}
where $\mathcal{I}_{H,2}(y)$ is defined in Eq.~(\ref{eq:IH2}). The chiral susceptibilities are
\begin{equation}
\begin{split}
\label{eq:chiu}
\chi_{u}(\theta,H)=&-\frac{B^{2}f^{2}}{\mathring{m}_{\pi}^{2}(\theta)}\left[1-\frac{B^{2}(m_{u}+m_{d}\cos\theta)^{2}}{\mathring{m}_{\pi}^{4}(\theta)}\right]-2B^{2}(l^{r}_{3}+h^{2}_{1}+h_{3})\\
&+\frac{3B^{2}}{2(4\pi)^{2}}\left[\frac{B^{2}(m_{u}+m_{d}\cos\theta)^{2}}{\mathring{m}_{\pi}^{4}(\theta)}-\log\frac{\Lambda^{2}}{\mathring{m}_{\pi}^{2}(\theta)}\right]\\
&-l_{7}\frac{8B^{4}m_{d}^{2}}{\mathring{m}_{\pi}^{4}(\theta)}\sin^{2}\theta\left[1-\frac{B^{2}m_{u}(5m_{u}+4m_{d}\cos\theta)}{\mathring{m}_{\pi}^{4}(\theta)}+\frac{4B^{4}m_{u}^{2}(m_{u}+m_{d}\cos\theta)^{2}}{\mathring{m}_{\pi}^{8}(\theta)}\right]\\
&+\frac{eH}{(4\pi)^{2}}\frac{B^{2}}{\mathring{m}_{\pi}^{2}(\theta)}\left[1-\frac{B^{2}(m_{u}+m_{d}\cos\theta)^{2}}{\mathring{m}_{\pi}^{4}(\theta)}\right]\mathcal{I}_{H,2}(\tfrac{\mathring{m}_{\pi}^{2}(\theta)}{eH})\\
&-\frac{1}{(4\pi)^{2}}\frac{B^{4}(m_{u}+m_{d}\cos\theta)^{2}}{\mathring{m}_{\pi}^{4}(\theta)}\mathcal{I}_{H,1}(\tfrac{\mathring{m}_{\pi}^{2}(\theta)}{eH})\ ,
\end{split}
\end{equation}
\begin{equation}
\begin{split}
\label{eq:chid}
\chi_{d}(\theta,H)=&-\frac{B^{2}f^{2}}{\mathring{m}_{\pi}^{2}(\theta)}\left[1-\frac{B^{2}(m_{d}+m_{u}\cos\theta)^{2}}{\mathring{m}_{\pi}^{4}(\theta)}\right]-2B^{2}(l^{r}_{3}+h^{2}_{1}+h_{3})\\
&+\frac{3B^{2}}{2(4\pi)^{2}}\left[\frac{B^{2}(m_{d}+m_{u}\cos\theta)^{2}}{\mathring{m}_{\pi}^{4}(\theta)}-\log\frac{\Lambda^{2}}{\mathring{m}_{\pi}^{2}(\theta)}\right]\\
&-l_{7}\frac{8B^{4}m_{u}^{2}}{\mathring{m}_{\pi}^{4}(\theta)}\sin^{2}\theta\left[1-\frac{B^{2}m_{d}(5m_{d}+4m_{u}\cos\theta)}{\mathring{m}_{\pi}^{4}(\theta)}+\frac{4B^{4}m_{d}^{2}(m_{d}+m_{u}\cos\theta)^{2}}{\mathring{m}_{\pi}^{8}(\theta)}\right]\\
&+\frac{eH}{(4\pi)^{2}}\frac{B^{2}}{\mathring{m}_{\pi}^{2}(\theta)}\left[1-\frac{B^{2}(m_{d}+m_{u}\cos\theta)^{2}}{\mathring{m}_{\pi}^{4}(\theta)}\right]\mathcal{I}_{H,2}(\tfrac{\mathring{m}_{\pi}^{2}(\theta)}{eH})\\
&-\frac{1}{(4\pi)^{2}}\frac{B^{2}(m_{d}+m_{u}\cos\theta)^{2}}{\mathring{m}_{\pi}^{2}(\theta)}\mathcal{I}_{H,1}(\tfrac{\mathring{m}_{\pi}^{2}(\theta)}{eH})\ ,
\end{split}
\end{equation}
where $\mathcal{I}_{H,1}(y)$ is defined in Eq.~(\ref{eq:IH1}). We note that the up and down quark condensates and chiral susceptibilities in the $\theta$-vacua are related by an interchange of the up and down quark masses.
\section{Discussion}
\label{sec:discussion}
\noindent
In this section, we characterize and discuss the relative shifts in the topological density, topological susceptibility and the fourth cumulant not only the $\theta$ vacuum and but also in the normal $\theta=0$ vacuum. In order to do, we begin by defining the relative shift of the quantity, $\mathcal{Q}$, as
\begin{equation}
\begin{split}
\label{eq:RO}
\mathcal{R}_{\mathcal{Q}}=\frac{\mathcal{Q}_{H}}{\mathcal{Q}_{\rm 0}}\ ,
\end{split}
\end{equation}
where $\mathcal{Q}=\mathcal{Q}_{0}+\mathcal{Q}_{H}$, with $\mathcal{Q}_{H}$ being the absolute shift of the quantity $\mathcal{Q}$ from its $H=0$ vacuum value, $\mathcal{Q}_{0}$, due to the background magnetic field at next-to-leading order or $\mathcal{O}(p^{4})$. Since we are interested in the leading order relative shifts and $\mathcal{Q}_{H}$ is of $\mathcal{O}(p^4)$, we can proceed by choosing $\mathcal{Q}_{0}$ to be of $\mathcal{O}(p^2)$, i.e. tree-level. 
\subsection{$\theta=0$}
\noindent
The relative shifts of the topological susceptibility and the fourth cumulant in the $\theta=0$ vacuum are easily found using Eqs.~(\ref{eq:chit}) and (\ref{eq:c4}),
\begin{align}
\label{eq:RchitRc4}
\mathcal{R}_{\chi_{t}}&=-\frac{eH}{(4\pi f)^{2}}\mathcal{I}_{H,2}(\tfrac{\mathring{m}_{\pi}^{2}(0)}{eH})\\
\mathcal{R}_{c_{4}}&=-\frac{eH}{(4\pi f)^{2}}\mathcal{I}_{H,2}(\tfrac{\mathring{m}_{\pi}^{2}(0)}{eH})+\frac{3}{(4\pi f)^{2}}\frac{Bm^{[3]}}{\bar{m}^{2}}\mathcal{I}_{H,1}(\tfrac{\mathring{m}_{\pi}^{2}(0)}{eH})\ ,
\end{align}
where $\mathring{m}_{\pi}(0)$ is the bare pion mass in the $\theta=0$ vacuum defined in Eq.~(\ref{eq:pionmass}), and the integrals $\mathcal{I}_{H,2}$ and $\mathcal{I}_{H,1}$ are defined in Eqs.~(\ref{eq:IH2}) and (\ref{eq:IH1}) respectively.  The reduced mass, $\bar{m}$, and the quantity, $m^{[3]}$, with mass dimension 3, are  defined as
\begin{equation}
\label{eq:masses}
\bar{m}=\left(\frac{1}{m_{u}}+\frac{1}{m_{d}}\right)^{-1}\ ,\  \ \ m^{[3]}=\left(\frac{1}{m_{u}^{3}}+\frac{1}{m_{d}^{3}}\right)^{-1}\ .
\end{equation}
\begin{figure}
	\centering
	\begin{subfigure}[b]{0.48\textwidth}
	\includegraphics[width=\textwidth]{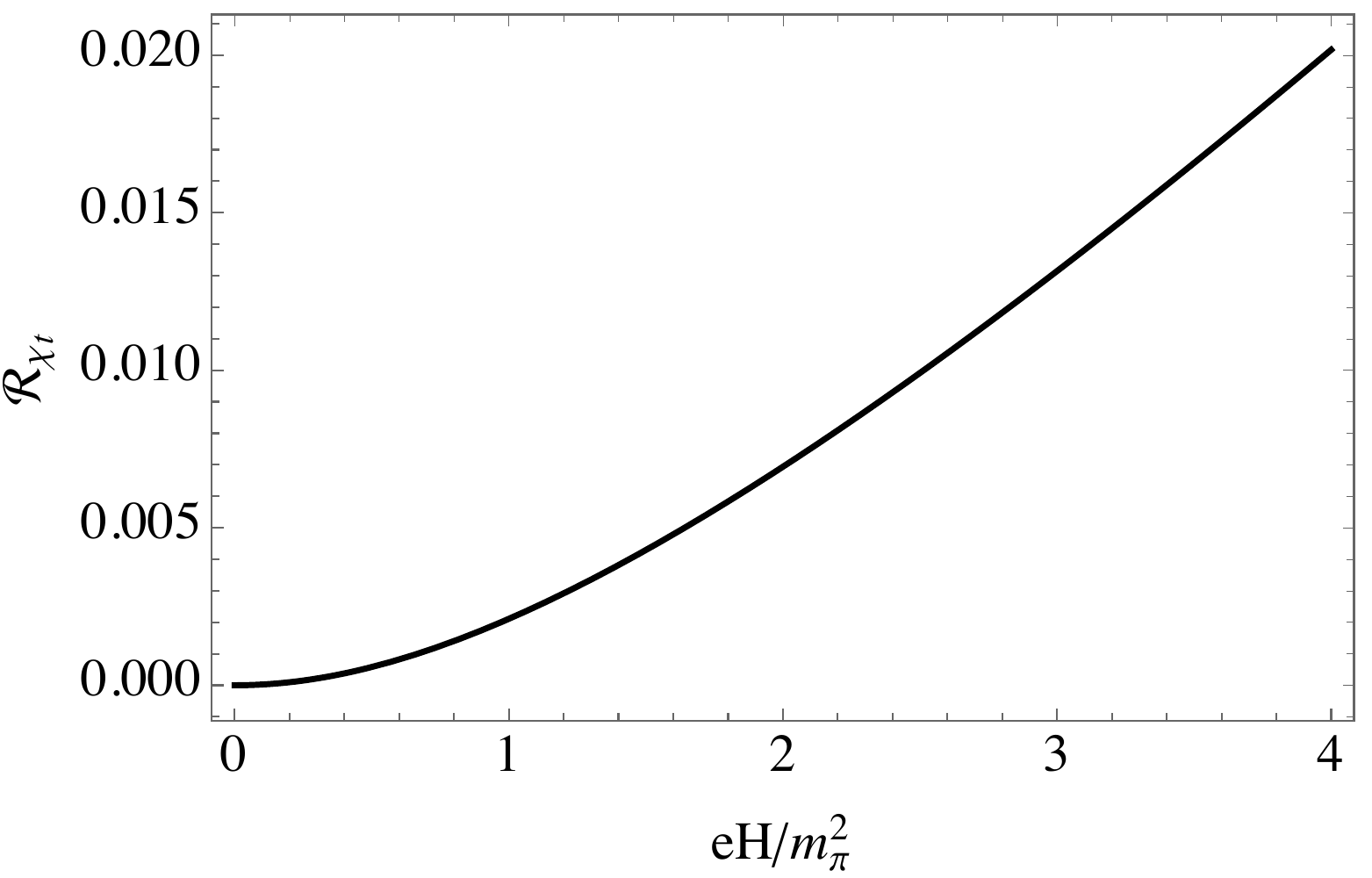}
	\end{subfigure}
	\begin{subfigure}[b]{0.48\textwidth}
	\includegraphics[width=\textwidth]{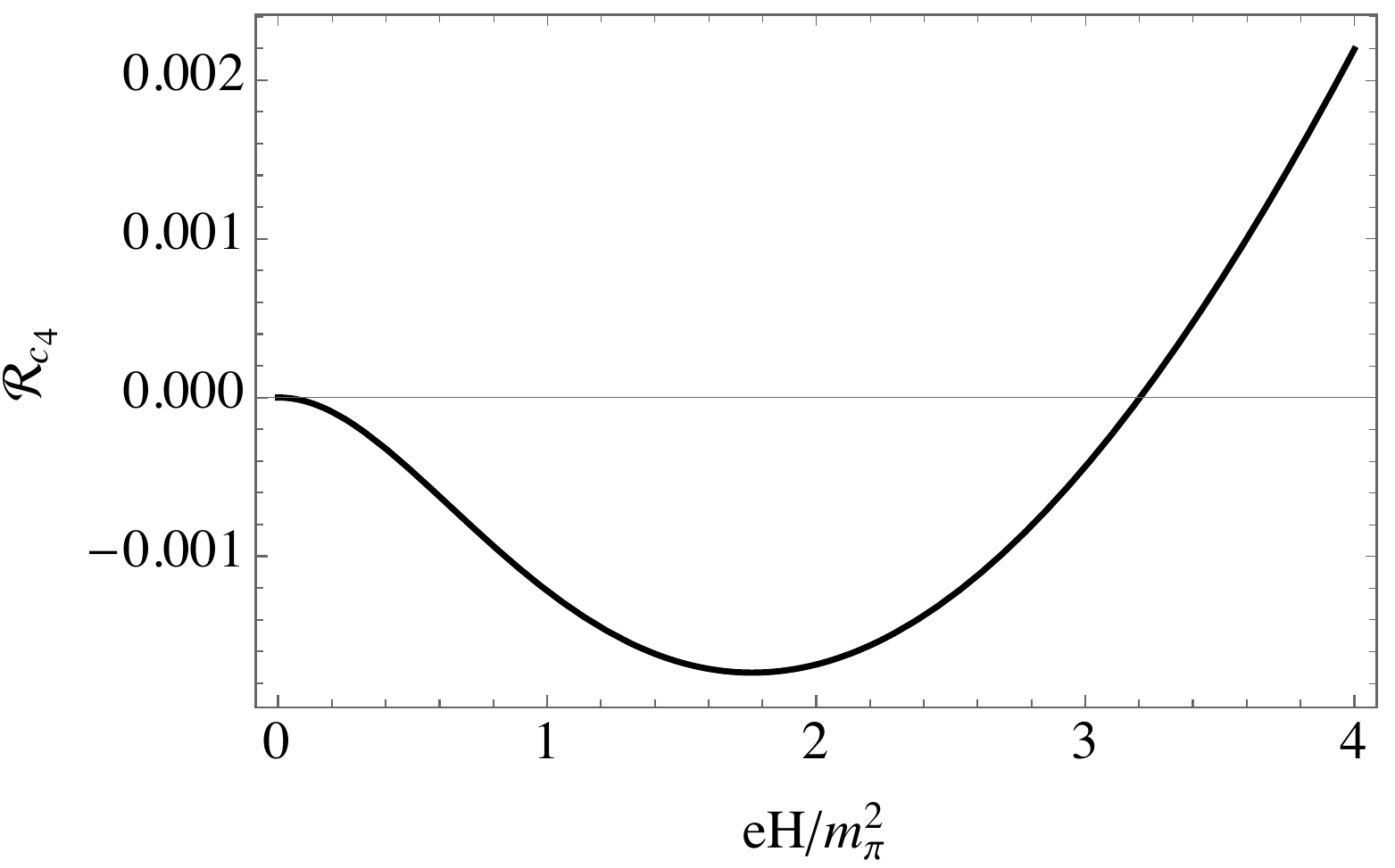}
	\end{subfigure}
\caption{Plots of the relative shift in the topological susceptibility (fourth cumulant) relative to their $H=0$ vacuum values as a function the background magnetic field, $eH$, on the left (right) panel.}
	\label{fig:rchitc4}
\end{figure}

In Fig.~\ref{fig:rchitc4}, we plot the relative shifts of the topological susceptibility and the fourth cumulant as a function of the external magnetic field in the $\theta=0$ vacuum. In order to do so, we replace the bare pion mass and decay constant in the relative shifts with the physical ones since the corrections that arise are an order higher than our calculations. Furthermore, we use the following values for the pion mass and decay constant extracted from the Particle Data Group review~\cite{Zyla:2020zbs}, 
\begin{equation}
\begin{split}
\label{eq:massPDG}
m_{\pi}=139.6\ {\rm MeV},\ \sqrt{2}f_{\pi}=130.2\ {\rm MeV},\ m_{u}=2.32\ {\rm MeV},\ m_{d}=3.71\ {\rm MeV}\ .
\end{split}
\end{equation} 
The relative shifts are plotted up to $eH=4m_{\pi}^{2}$, which is within the regime of validity of $\chi$PT. 

As can be seen in the left panel of Fig.~\ref{fig:rchitc4}, the topological susceptibility increases monotically with a change of approximately $2\%$ for a magnetic field, $eH=4m_{\pi}^{2}$. The monotonic increase is anticipated from Eq.~(\ref{eq:RchitRc4}) since the integrals, $I_{H,n}$, are negative definite, vanishing in the absence of a background field. 

In the right panel of Fig.~\ref{fig:rchitc4}, we plot the relative shift of the fourth cumulant. It is suppressed at low magnetic fields up to approximately $eH\approx3.2m_{\pi}^{2}$ and is monotonically enhanced thereafter. This can be anticipated from Eq.~(\ref{eq:RchitRc4}) since the two terms come with opposite signs with the second term dominating for weak magnetic fields and the first term for larger ones. It is also worth noting that the relative shift of the topological susceptibility has a structure very similar to that of the well-known relative shift of the light quark condensates in a uniform magnetic field, which has the form
\begin{equation}
\begin{split}
\mathcal{R}_{\langle\bar{q}_{f}q_{f}\rangle}&=-\frac{eH}{(4\pi f)^{2}}\mathcal{I}_{H,2}(\tfrac{\mathring{m}_{\pi}^{2}(0)}{eH})\ ,
\end{split}
\end{equation}
where $q_{f}=u$ or $d$, as can be deduced easily from Eqs.~(\ref{eq:uu}) and (\ref{eq:dd}). The relative shift of the topological susceptibility in Eq.~(\ref{eq:RchitRc4}) is precisely equal to that of the quark condensates. As we note in the following subsection, this is a manifestation of a sum rule whereby the topological susceptibility shift is proportional to the quark condensate shifts.
\subsubsection{Sum Rules ($\theta=0$)}

There are sum rules that connect the topological susceptibility shift to the quark condensate shifts and the fourth cumulant shift to the shifts of the quark condensates and quark susceptibilities~{\cite{Adhikari:2021lbl}}. 

We first proceed by using Eqs.~(\ref{eq:chit}) and (\ref{eq:c4}) to find the absolute shifts  of the topological susceptibility and the fourth cumulant, which gives
\begin{align}
\chi_{t,H}&=-\frac{B\bar{m}}{(4\pi)^{2}}(eH)\mathcal{I}_{H,2}(\tfrac{\mathring{m}_{\pi}^{2}(0)}{eH})\\
c_{4,H}&=\frac{B\bar{m}^{4}}{(4\pi)^{2}}\left(\frac{eH}{m^{[3]}}\right)\mathcal{I}_{H,2}(\tfrac{\mathring{m}_{\pi}^{2}(0)}{eH})-\frac{3B^{2}\bar{m}^{2}}{(4\pi)^{2}}\mathcal{I}_{H,1}(\tfrac{\mathring{m}_{\pi}^{2}(0)}{eH})\ ,
\end{align}
where $\bar{m}$ is the reduced mass defined in Eq. (\ref{eq:masses}) and $\mathring{m}_{\pi}(0)$ is the bare pion mass in the $\theta=0$ vacuum. Similarly, using Eqs. (\ref{eq:uu}), (\ref{eq:dd}), (\ref{eq:chiu}), we get for the up quark and down quark condensates and susceptibilities,
\begin{align}
\langle\bar{q}_{f}q_{f}\rangle_{H}=\frac{B(eH)}{(4\pi)^{2}}\mathcal{I}_{H,2}(\tfrac{\mathring{m}_{\pi}^{2}(0)}{eH})\ ,\ \chi_{q_{f},H}=-\frac{B^{2}}{(4\pi)^{2}}\mathcal{I}_{H,1}(\tfrac{\mathring{m}_{\pi}^{2}(0)}{eH})\ ,
\end{align}
where $q_{f}$ is either $u$ or $d$. Since both the topological shift and quark condensate shifts are proportional to the integral $\mathcal{I}_{H,2}$ we can straightforwardly relate these shifts. We find 
\begin{align}
\chi_{t,H}&=-\bar{m}\langle\bar{q}_{f}q_{f}\rangle_{H}\ ,
\end{align}
which shows that the topological susceptibility shift is directly proportional to the chiral condensate shift and the shifts have opposite signs. While the quark condensates become more negative in the presence of a magnetic field, i.e. magnetic catalysis, the topological susceptibility is enhanced as we have previously observed. 

Similarly, the shift in the fourth cumulant is proportional to $\mathcal{I}_{n,2}$ and $\mathcal{I}_{n,1}$ and consequently can be related to the quark condensate shifts, which are proportional to $\mathcal{I}_{n,2}$ and the quark susceptibility shifts, which are proportional to $\mathcal{I}_{n,1}$. We get the following sum rule,
\begin{equation}
\begin{split}
c_{4,H}&=\bar{m}^{4}\left(\sum_{q_{f}=u,d}\frac{\langle\bar{q}_{f}q_{f}\rangle_{H}}{m^{3}_{q_{f}}}\right)+3\bar{m}^{2}\chi_{q_{f},H}\ ,
\end{split}
\end{equation}
where $\bar{m}$ is the reduced mass defined in Eq.~(\ref{eq:masses}). As anticipated, the shift in the fourth cumulant is proportional to both the quark condensate shifts and the chiral susceptibility shifts.

Interestingly, the sum rule for the topological susceptibility shift is a manifestation of a Ward-Takahashi identity that relates the topological susceptibility shift to the shift of the quark condensates and the shift of an integrated two-point correlation function of $\bar{q}_{f}\gamma_{5}q_{f}$, which is proportional to the squares of the quark masses, and is consequently suppressed for small quark masses relative to the term linear in the quark mass, which is also proportional to the quark condensate. The exact result can be derived using the QCD partition function with the $\theta$-term. We refer the interested reader for a full discussion in Ref.~\cite{Adhikari:2021lbl}. 
\subsection{$\theta\neq 0$}
\begin{figure}
	\centering
	\includegraphics[width=.48\textwidth]{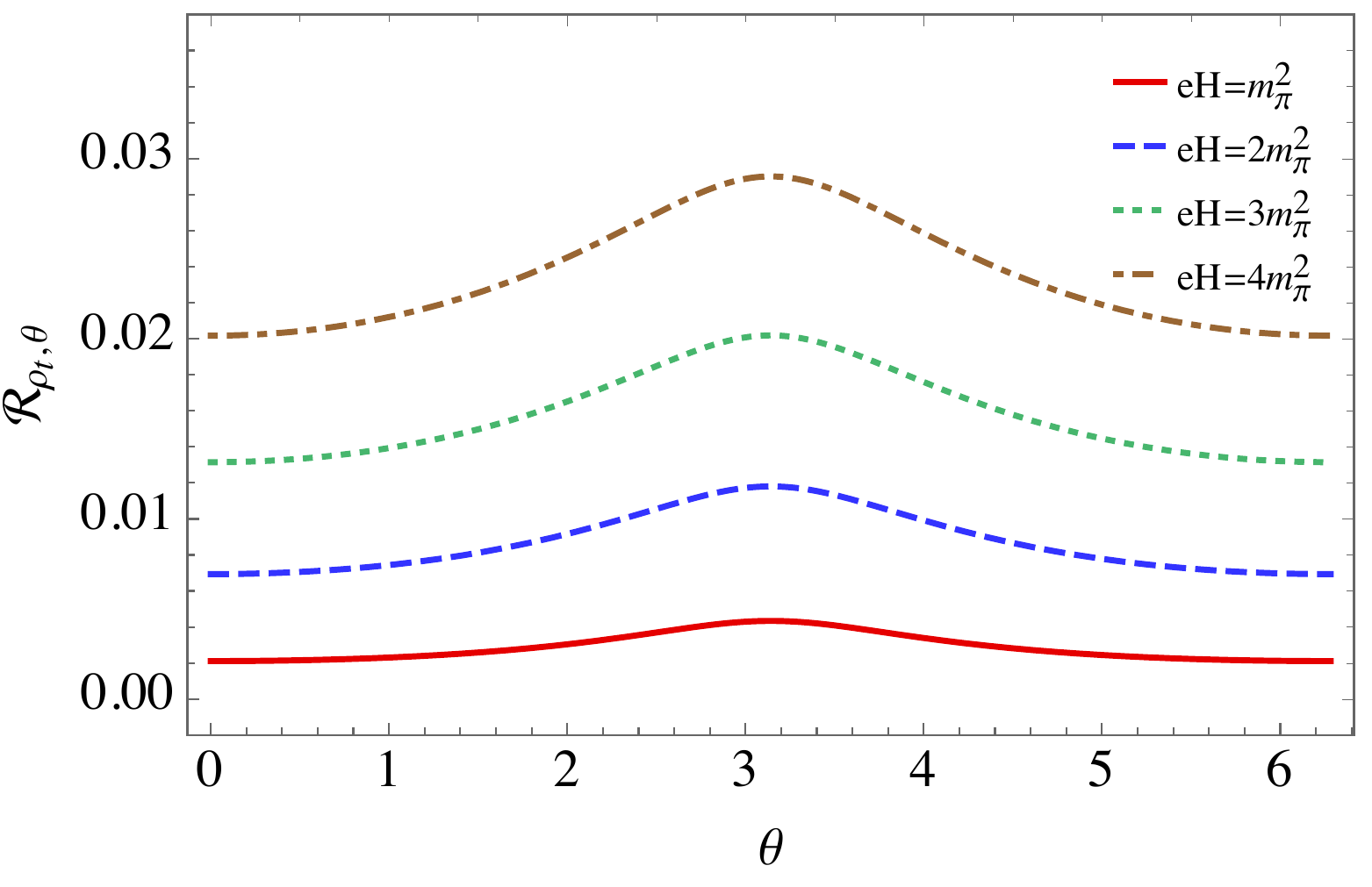}
\caption{Plot of the relative shift of the topological density as a function of the magnetic field, $eH$.}
	\label{fig:Rrhot}
\end{figure}
In the $\theta$-vacuum of QCD, i.e. $0<\theta<2\pi$, CP-odd quantities, which are zero when $\theta=0$ become finite. For instance, the topological density calculated in Eq.~(\ref{eq:topden}) is proportional to $\sin\theta$, which means it is an odd function about $\theta=\pi$, and possesses the following relative shift (compared to its tree-level value)
\begin{equation}
\begin{split}
\label{eq:Rrhot}
\mathcal{R}_{\rho_{t}}&=-\frac{(eH)^{2}}{(4\pi f)^{2}}\mathcal{I}_{H,2}(\tfrac{\mathring{m}_{\pi}^{2}(\theta)}{eH})\ .
\end{split}
\end{equation}
The structure is quite similar to that of the topological susceptibility shift in Eq.~(\ref{eq:RchitRc4}) with the bare pion mass here belonging to $\theta$ vacuum. (It is also important to note that the relative shift relationship does not apply for $\theta\mod2\pi=0$ and $\pi$ since we would be dividing by zero.) Since $\mathcal{I}_{H,2}$ is negative for finite values of $H$, the topological density is enhanced (monotonically) by the magnetic field with the relative shift being symmetric (even) about $\theta=\pi$. In Fig.~\ref{fig:Rrhot}, we plot the relative shift of the topological density in the $\theta$-vacuum for different values of the magnetic field ($eH$). The magnitude of the topological density is enhanced for each value of $\theta$. For each value of the magnetic field, the topological density first increases from $0$ to $\pi$, then decreases from $\pi$ to $2\pi$. This is due to Eq.~(\ref{eq:Rrhot}) and the fact that the pion mass in the $\theta$-vacuum decreases from $0$ to $\pi$ and then decreases from $\pi$ to $2\pi$ as can be seen in the right panel of Fig.~\ref{fig:gs-pionmass}.

\begin{figure}
	\centering
	\begin{subfigure}[b]{0.48\textwidth}
	\includegraphics[width=\textwidth]{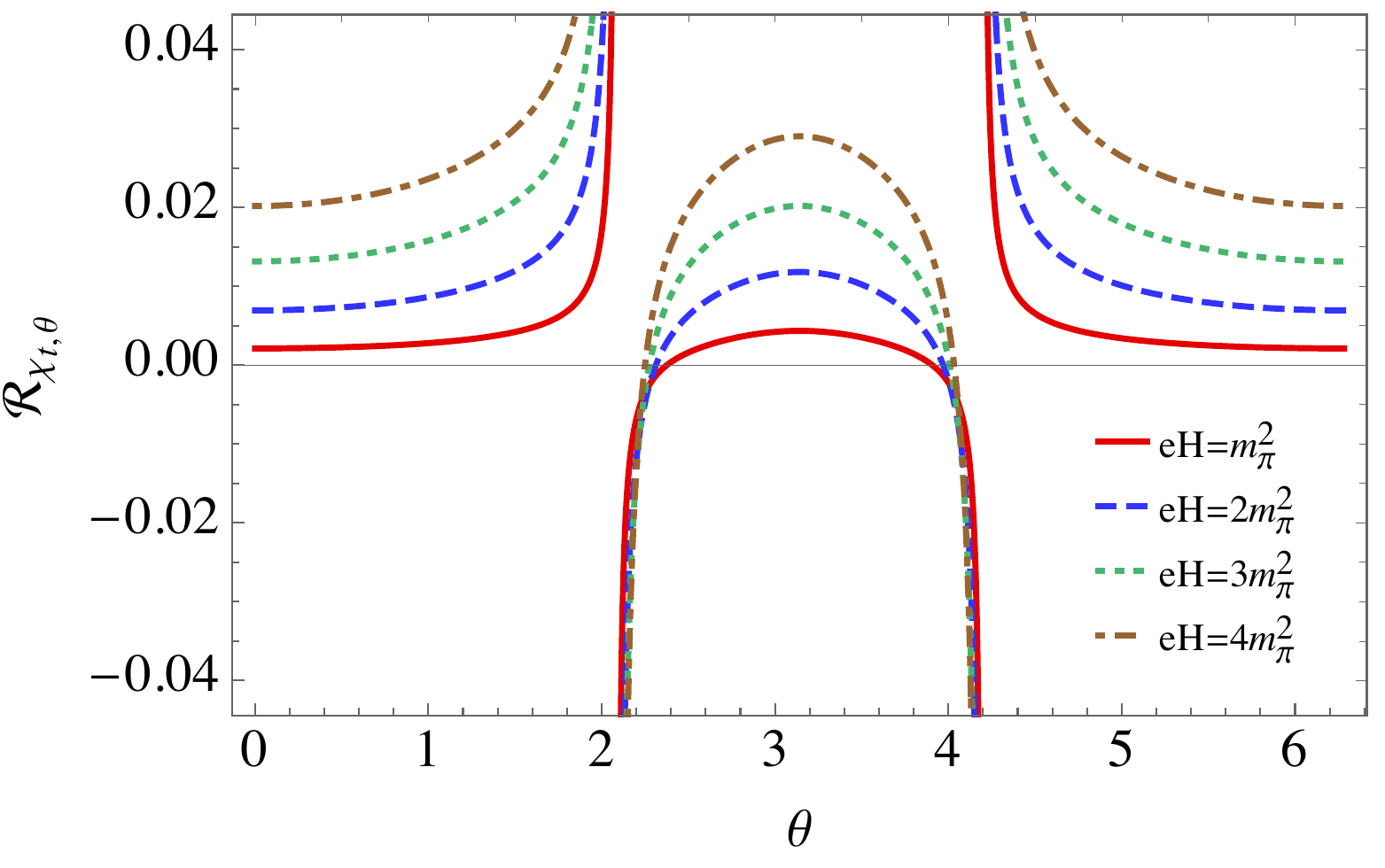}
	\end{subfigure}
	\begin{subfigure}[b]{0.48\textwidth}
	\includegraphics[width=\textwidth]{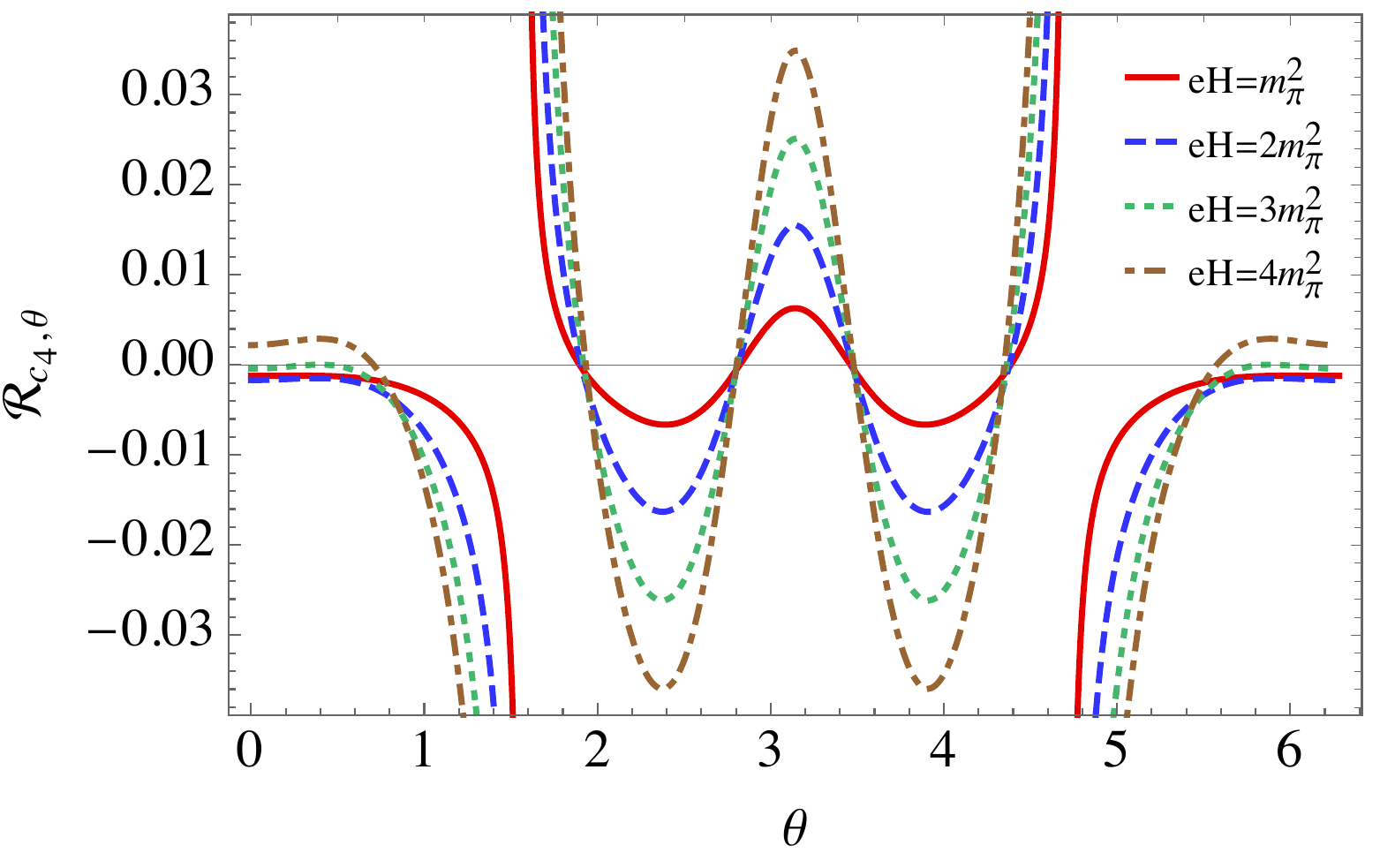}
	\end{subfigure}
\caption{Plots of the relative shift in the topological susceptibility (left panel) and the relative shift of the fourth cumulant (right panel) as a function of the magnetic field, $eH$.}
	\label{fig:rchithetac4theta}
\end{figure}

In Fig.~\ref{fig:rchithetac4theta}, we plot the relative shifts in the topological susceptibility and the fourth cumulant relative to their tree-level, $\theta$-dependent values for four different values of the magnetic field $\tfrac{eH}{m_{\pi}^{2}}=1.0,2.0,3.0,4.0$. These are denoted by red (solid), blue (dashed), green (dotted) and brown (dot-dashed) lines. Using Eqs.~(\ref{eq:chit}) and (\ref{eq:RO}), the relative shift of the topological susceptibility is
\begin{equation}
\begin{split}
\label{eq:Rchittheta}
\mathcal{R}_{\chi_{t},\theta}&=-\frac{eH}{(4\pi f)^{2}}\mathcal{I}_{H,2}(\tfrac{\mathring{m}_{\pi}^{2}(\theta)}{eH})-\frac{1}{(4\pi f)^{2}}\frac{\left(\frac{B^{2}m_{u}m_{d}}{\mathring{m}_{\pi}^{2}(\theta)}\sin^{2}\theta\right)}{\left(\cos\theta+\frac{B^{2}m_{u}m_{d}}{\mathring{m}_{\pi}^{4}(\theta)}\sin^{2}\theta\right)}\mathcal{I}_{H,1}(\tfrac{\mathring{m}_{\pi}^{2}(\theta)}{eH})\ ,
\end{split}
\end{equation}
where we note that for $\theta=0$, the result reduces to that of Eq.~(\ref{eq:RchitRc4}), unlike which it also contains a term proportional to $\mathcal{I}_{H,1}$, which becomes divergent for two values of $\theta$ between $0$ and $2\pi$. These divergences arise from the topological susceptibility at tree-level, i.e. the first term of Eq.~(\ref{eq:chit}), which are the values at which the tree-level topological susceptibility changes sign. As such in the left panel of Fig.~\ref{fig:rchithetac4theta}, we observe two points for each magnetic field where the relative shifts diverge. The divergences shift from the tree-level values at finite $H$ due to the additive, first term in Eq.~(\ref{eq:Rchittheta}). The relative shift is positive and increases with increasing magnetic field except for a narrow range of values of $\theta$ near the zeroes of the tree-level susceptibility. 

In the right panel of Fig.~\ref{fig:rchithetac4theta}, we plot the relative shift of the fourth cumulant, where the absolute shift is given by the last four terms of Eq.~(\ref{eq:c4}) (which are proportional to $\mathcal{I}_{H,n}$) and the tree-level fourth cumulant is the first term. We forgo the explicit writing of the relative shift since it is cumbersome and not very informative. The tree-level fourth cumulant is negative with zeroes at two different values of $\theta$. As such we notice divergences in the relative shift of the fourth cumulant in the right panel of Fig.~\ref{fig:rchithetac4theta}. The magnetic field both enhances and diminishes the magnitude of the fourth cumulant: in the region near the zeroes of its tree-level value, the fourth cumulant is either monotonically enhanced and suppressed. This behavior is different from the behavior at $\theta=0$, where the fourth cumulant is suppressed for small magnetic fields and enhanced for large fields as previously seen in the right panel of Fig.~\ref{fig:rchitc4}. This behavior holds for $\theta\ge0$ and approximately less than $0.75$ (and similarly for $\theta\le 2\pi$ and approximately greater than $2\pi$ minus $0.75$).
\section{Summary}
With the discussion complete, we summarize and present some outlook on the key findings of the study conducted in this paper: we have characterized the topological density, the topological susceptibility and fourth cumulant of the QCD $\theta$ vacuum using a model-independent calculation in $\chi$PT performed in the simultaneous presence of both $\theta$ and a background magnetic field. Our results are valid for weak fields and small masses within the regime of validity of $\chi$PT. In the simultaneous presence of the QCD $\theta$ angle and a magnetic field, QCD suffers from the fermion sign problem -- as such the results presented here are not amenable to lattice studies unlike at $\theta=0$ -- see Ref.~\cite{VICARI200993} for a discussion of previous lattice studies at $H=0$. Furthermore, in this study, we have identified sum rules (valid for $\theta=0$), first noted in Ref.~\cite{Adhikari:2021lbl}, for completeness and also found a new sum rule that the relative shift of the topological density is equal to that of the up and down quark condensates. Finally, we also found that the magnetic field enhances the size of the topological susceptibility (except for narrow values of $\theta$) while it can either diminish or enhance the fourth cumulant with the result depending on the values of $\theta$ and the magnetic field. The topological density, on the other hand, if non-zero in the $\theta$-vacuum is monotonically enhanced for all magnetic fields within the regime of validity of $\chi$PT. 
\section*{Acknowledgements} 
\noindent 
P.A. would like to acknowledge the support of St. Olaf College start up funds.
\appendix
\section{Useful Integrals}
\label{integrals}
\noindent
The finite part of the one-loop effective potential of a charged particle in Eq.~(\ref{eq:IHfin}) is most conveniently characterized as
\begin{equation}
\begin{split}
\label{Schwingerint}
I_{H}^{\rm fin}(m)&=\frac{(eH)^{2}}{(4\pi)^{2}}\mathfrak{I}_{H}(\tfrac{m^{2}}{eH})\\
\mathfrak{I}_{H}(y)&=-\int_{0}^{\infty}dz\ \frac{e^{-yz}}{z^{3}}\left[\frac{z}{\sinh z}-1+\frac{z^{2}}{6}\right]\\
&=4\zeta^{(1,0)}(-1,\tfrac{y+1}{2})+(\tfrac{y}{2})^{2}(1-2\log \tfrac{y}{2})+\tfrac{1}{6}(\log \tfrac{y}{2}+1)\ ,
\end{split}
\end{equation}
 where $y\equiv\frac{m^{2}}{eH}$ and $\zeta(s,a)$ is the Hurwitz zeta function. In characterizing the topological susceptibility and the fourth cumulant the following dimensionless integrals are useful,
\begin{equation}
\begin{split}
\label{eq:IHn}
\mathcal{I}_{H,n}(y)&=\int_{0}^{\infty}dz\frac{e^{-yz}}{z^{n}}\left(\frac{z}{\sinh z}-1\right)\ .
\end{split}
\end{equation}
in particular for $n=2,1,0,$ and $-1$. They are
\begin{align}
\label{eq:IH2}
\mathcal{I}_{H,2}(y)&=2\zeta^{(1,0)}(0,\tfrac{y+1}{2})-y(\log \tfrac{y}{2}-1)\\
\label{eq:IH1}
\mathcal{I}_{H,1}(y)&=\log\tfrac{y}{2}-\psi_{0}\left (\tfrac{y+1}{2} \right)\\
\label{eq:IH0}
\mathcal{I}_{H,0}(y)&=-\tfrac{1}{y}+\tfrac{1}{2}\psi_{1}(\tfrac{y+1}{2})\\
\label{eq:IH-1}
\mathcal{I}_{H,-1}(y)&=-\tfrac{1}{y^{2}}-\tfrac{1}{4}\psi_{2}(\tfrac{y+1}{2})\ ,
\end{align}
where $\zeta(s,a)$ is the Hurwitz zeta function with the two superscripts on the function representing derivatives with respect to the first variable $s$ and the second variable $a$ respectively, $\psi_{n}(y)$ is the polygamma function, which is defined as the derivative of the log of the gamma function: $\psi_{n}(y)=\frac{d^{n+1}\log\Gamma(y)}{dy^{n+1}}$. It is also worth noting that the integrals, $\mathcal{I}_{H,n}$ are negative definite and asymptote to zero as $y\rightarrow \infty$, i.e. vanish in the absence of a magnetic field. Finally, $\mathcal{I}_{H,n}$ satisfy the following simple derivative identity 
\begin{align}
\frac{\partial \mathcal{I}_{H,n}(y)}{\partial y}=-\mathcal{I}_{H,n-1}(y)\ ,
\end{align}
where $y$ is a dimensionless ratio, $y\equiv\frac{m^{2}}{eH}$.
\bibliographystyle{apsrev4-1}
\bibliography{/Users/prabal7e7/Documents/Research/bib}
\end{document}